\documentclass[12pt,epsf]{article}
\usepackage{amsmath}
\usepackage{amsthm, amssymb}
\usepackage{graphicx}
\usepackage{setspace}
\usepackage{subfigure}
\usepackage{cite}

\theoremstyle{remark}

\newcommand{\be}{\begin{equation}}
\newcommand{\ee}{\end{equation}}
\newcommand{\bea}{\begin{eqnarray}}
\newcommand{\eea}{\end{eqnarray}}
\newcommand{\bear}{\begin{eqnarray}}
\newcommand{\eear}{\end{eqnarray}}
\newcommand{\beas}{\begin{eqnarray*}}

\newcommand{\eeas}{\end{eqnarray*}}
\newcommand{\ba}{\begin{array}}
\newcommand{\ea}{\end{array}}

\newcommand{\ra}

\newcommand{\pd}[2][1]{\ifnum#1=1 \frac{\partial}{\partial {#2}} \else
  \frac{\partial^#1}{\partial {#2}^{#1}}\fi}
\newcommand{\dpd}[2][1]{\ifnum#1=1 \dfrac{\partial}{\partial {#2}} \else
  \frac{\partial^#1}{\partial {#2}^{#1}}\fi}
\newcommand{\td}[2][1]{\ifnum#1=1 \frac{d}{d{#2}} \else
  \frac{d^#1}{d{#2}^{#1}}\fi}

\newcommand{\nbox}{{\,\lower0.9pt\vbox{\hrule \hbox{\vrule height 0.2 cm \hskip 0.19 cm \vrule height 0.2 cm}\hrule}\,}}

\def\href#1#2{#2}

\usepackage{color}

\usepackage{xcolor}
\usepackage[citebordercolor=green, linkbordercolor={ 0 0 1}, linktocpage=true]{hyperref}

\newcommand\blfootnote[1]{%
  \begingroup
  \renewcommand\thefootnote{}\footnote{#1}%
  \addtocounter{footnote}{-1}%
  \endgroup
}

\begin{document}
\begin{titlepage}
\begin{NoHyper}
\hfill
\vbox{
    \halign{#\hfil         \cr
           } % end of \halign
      }  % end of \vbox
\vspace*{20mm}
\begin{center}
{\Large \bf Into the Bulk: A Covariant Approach}

\vspace*{15mm}
\vspace*{1mm}
Netta Engelhardt
\vspace*{1cm}
\blfootnote{nengelhardt@princeton.edu}

{Department of Physics, Princeton University\\
Princeton, NJ 08544 USA}

\vspace*{1cm}
%%\maketitle
\end{center}
\begin{abstract}
I propose a general, covariant way of defining when one region is ``deeper in the bulk'' than another.  This definition is formulated outside of an event horizon (or in the absence thereof) in generic geometries; it may be applied to both points and surfaces, and may be used to compare the depth of bulk points or surfaces relative to a particular boundary subregion or relative to the entire boundary. Using the recently proposed ``lightcone cut'' formalism, the comparative depth between two bulk points can be determined from the singularity structure of Lorentzian correlators in the dual field theory. I prove that, by this definition, causal wedges of progressively larger regions probe monotonically deeper in the bulk. The definition furthermore matches expectations in pure AdS and in static AdS black holes with isotropic spatial slices, where a well-defined holographic coordinate exists. In terms of holographic RG flow, this new definition of bulk depth makes contact with coarse-graining over both large distances and long time scales. 

\end{abstract}
\end{NoHyper}

\end{titlepage}
\tableofcontents
\vskip 1cm
\begin{spacing}{1.2}
\section{Introduction}\label{sec:intro}

One of the more mysterious aspects of the celebrated AdS/CFT correspondence~\cite{Mal97, Wit98a, GubKle98} is the emergence of the holographic dimension. AdS/CFT, a particular realization of the holographic principle~\cite{Sus95, Tho93, Bou02}, posits that the dynamics of gravitational theories in a $(d+1)$-dimensional asymptotically Anti-de Sitter (AdS) spacetime can be described by a non-gravitational quantum field theory (QFT) in $d$ dimensions\footnote{The QFT is defined on a representative of the conformal class of the AdS boundary.}. The name \textit{holography} is itself derived from the equivalence of a lower-dimensional theory to a higher-dimensional one, although a precise understanding of the way in which the additional holographic dimension is described in the dual theory remains elusive.

Energy scale, and in particular renormalization group (RG) flow, has been suggested as the responsible party for the emergent holographic dimension (see e.g.~\cite{SusWit98, PeePol98, Akh98, BalKra99, FreGub99, DebVer99, deBo01, Lee09, FauLiu10, HeePol10}). Under such a hypothesis, greater depth in the bulk corresponds to coarse-graining in the QFT. Of particular note is the UV/IR correspondence: a UV cutoff $\Lambda$ in the field theory is dual in the bulk to a large ``radial'' cutoff at $r=\Lambda$.

More generally, the expectation is that geometry deep in the bulk should in some sense be dual to to the infrared physics of the dual field theory. Deep bulk geometry is usually reached by nonlocal observables across large distances on the boundary (e.g. the entanglement entropy of progressively larger boundary intervals is understood to probe deeper in the bulk via the Ryu-Takayanagi prescription~\cite{RyuTak06, HubRan07}), in agreement with the expectation that points deep in the bulk are sensitive to dual infrared energy scales. A clear example may be constructed from holographic field theories with a confinement/deconfinement phase transition (see e.g.~\cite{RyuTak06-2, NisTak06, KleKut07}). 
%with a mass gap, in which extremal surfaces dual to entanglement entropy manifest a topology change when dual to the IR~\cite{}.

At this point, a natural question arises:  how is ``bulk depth'' defined? We shall be primarily concerned with the comparative bulk depth between two points, but in order to make contact with holographic RG flow, we will give a definition that also encompasses the depths of surfaces.

A general asymptotically AdS spacetime does not have a natural holographic coordinate. Even the Fefferman-Graham expansion~\cite{FefGra85} fails to provide a unique coordinate; additionally, the Fefferman-Graham expansion often fails to converge away from the asymptotic region. Furthermore, for those spacetimes in which the Fefferman-Graham gauge is in fact well-defined everywhere, a coordinate-based definition is inherently unsatisfactory as it is not only gauge-dependent but also not naturally accessible from dual QFT data.

\begin{figure}[t!]
\centering
\includegraphics[width=7cm]{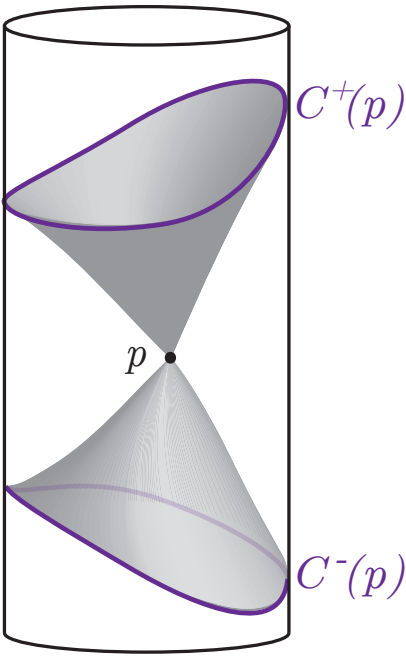}
\caption{The lightcone cuts $C^{\pm}(p)$ of a point $p$ in some generic asymptotically AdS geometry. The cuts are given by the intersection of the boundary of the past and future of $p$ with the asymptotic boundary; this is simply the intersection of the lightcone of $p$ with the asymptotic boundary, with generators leaving the surface after intersections. The irregular shape of the the cones serves to illiustrate the generic effects of gravitational lensing on the lightcone of a point.}
\label{fig:LCcuts}
\end{figure} 

Progress was made recently towards a qualification of bulk depth perception in~\cite{CzeLamMcC15a, CzeLam16a, daCGui16, LewTur16} (see also~\cite{Sah00a} for earlier work in spacetimes with timelike Killing symmetry). The approach of~\cite{CzeLamMcC15a, CzeLam16a} invokes the inverse Radon transform; as the inverse Radon transform is known only in AdS, we will pursue here a different line of investigation. Our goal is to give a covariant definition of bulk depth with minimal assumptions about the geometry. The crux of our construction is causality, which has been previously suggested as a key ingredient in the UV/IR correspondence~\cite{Sah00a, Bou09}.

\begin{figure}[t]
\centering
\includegraphics[width=7cm]{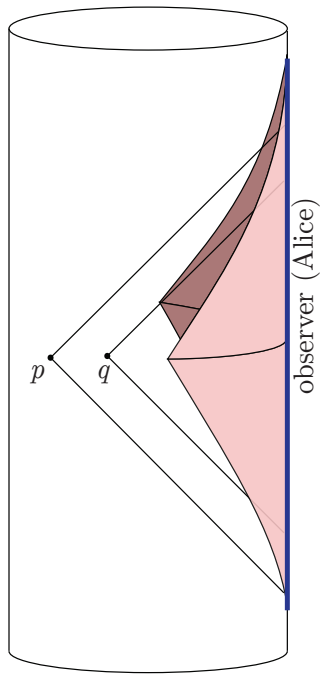}
\caption{A schematic illustrating an observer Alice (thick blue line) propagating on the causal diamond of some boundary region (pink). The points $p$ and $q$ live in the bulk; Alice first loses causal contact with $p$, then with $q$. She regains causal contact with $q$ before regaining causal contact with $p$: Alice perceives $p$ as farther away than $q$.}
\label{fig:intuition}
\end{figure} 

At a fundamental level, the question of comparing the depth of two bulk points is itself ill-posed: the specification of bulk points is not  \textit{per se} well-defined from field theory data. Fortunately, this potential pitfall has been addressed in several different ways in the literature in the large $N$, large $\lambda$ limit (see e.g.~\cite{HeeMar, Hee12, KabLif12, KabLif13, KabLif13b, CzeLam, MalSimZhi}, and most recently in the context of quantum error correction~\cite{AlmDon15, MinPol15}). We will take the approach of~\cite{MalSimZhi} (and earlier work~\cite{PolSus99, GarGid09, HeePen09, Pen10, OkuPen11}), and in particular its application in~\cite{EngHor16a}: a bulk point can be specified in terms of the intersection of the boundary of its past and future with the asymptotic boundary. The intersection of the past (future) lightcone of a point $p$ with the asymptotic boundary is called the past (future) \textit{lightcone cut} of $p$~\cite{New76, EngHor16a}; this is illustrated in Fig.~\ref{fig:LCcuts}. Any bulk point with both past and future causal contact with the boundary may thus be uniquely identified via the location of its lightcone cuts, which can themselves be obtained from the singularities of time-ordered Lorentzian correlators in the dual field theory~\cite{MalSimZhi}. This procedure will be reviewed below in Sec.~\ref{sec:rev}.

Having settled on a covariant definition of a bulk point from field theory data (at least for points within the causal wedge of the boundary), we turn to obtaining a general, covariant qualification of comparative bulk depth between two points.
%return to the main subject of this paper: qualifying depth perception of bulk points. 
Consider the following gedanken experiment. Let Alice be a boundary observer; intuitively, Alice should perceive a point $p$ to be ``deeper in the bulk'' than a point $q$ if she finds that she is out of causal contact with $p$ for a longer proper time than with $q$. See Fig.~\ref{fig:intuition}. In particular, if, in the time that it takes Alice to send and receive a null curve from $p$, she can send and receive timelike curves from $q$, then intuitively $q$ is ``closer'' to Alice than $p$.  In other words, the subset of a boundary causal diamond that is spacelike separated from $p$ is properly contained in the subset of the same causal diamond that is spacelike-separated from $q$. 

\begin{figure}[t!]
\centering
\subfigure[]{
\includegraphics[width=6cm]{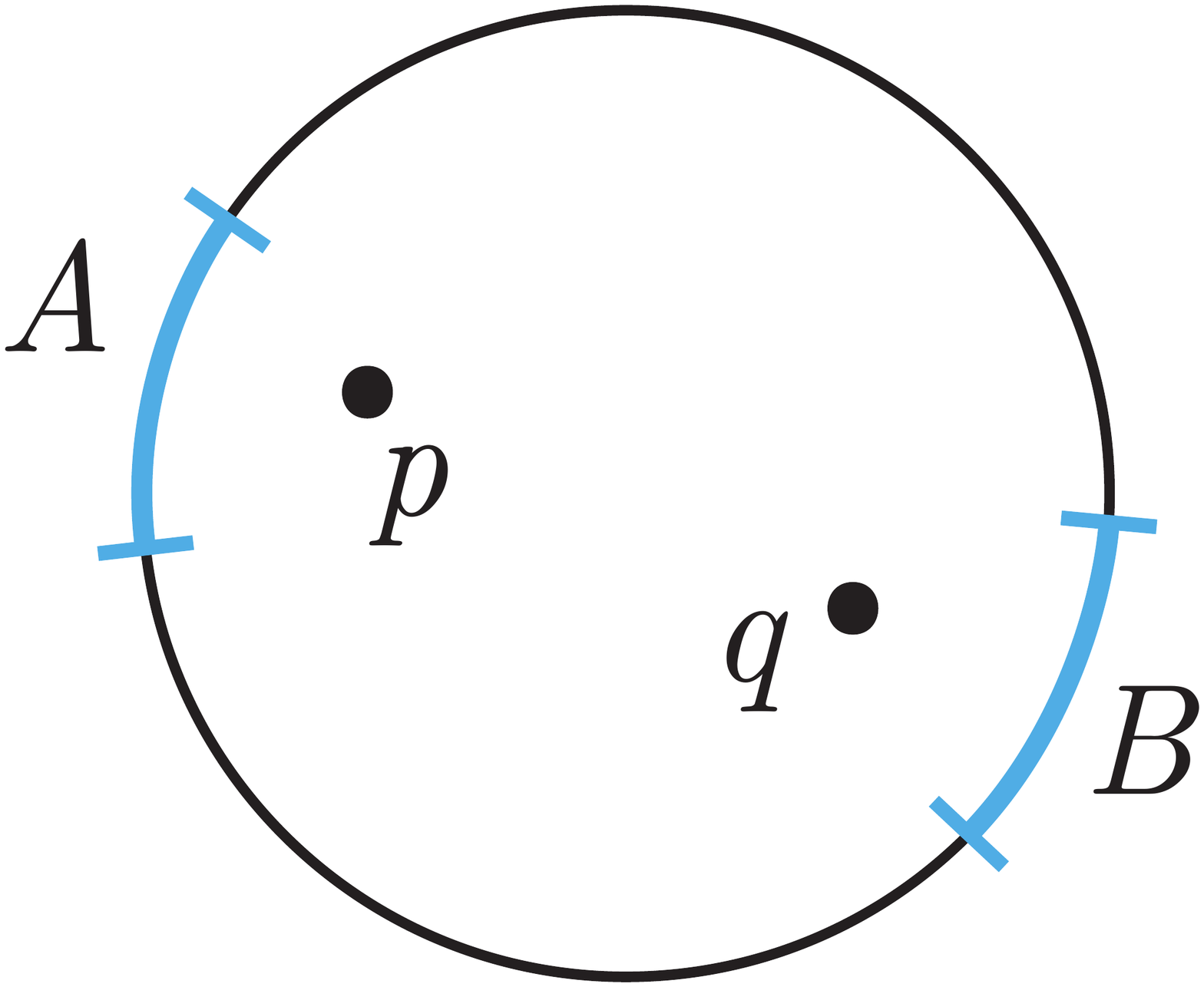}
\label{subfig:IntroAdSSlice}
}
\hspace{1cm}
\subfigure[]{
\includegraphics[width=6cm]{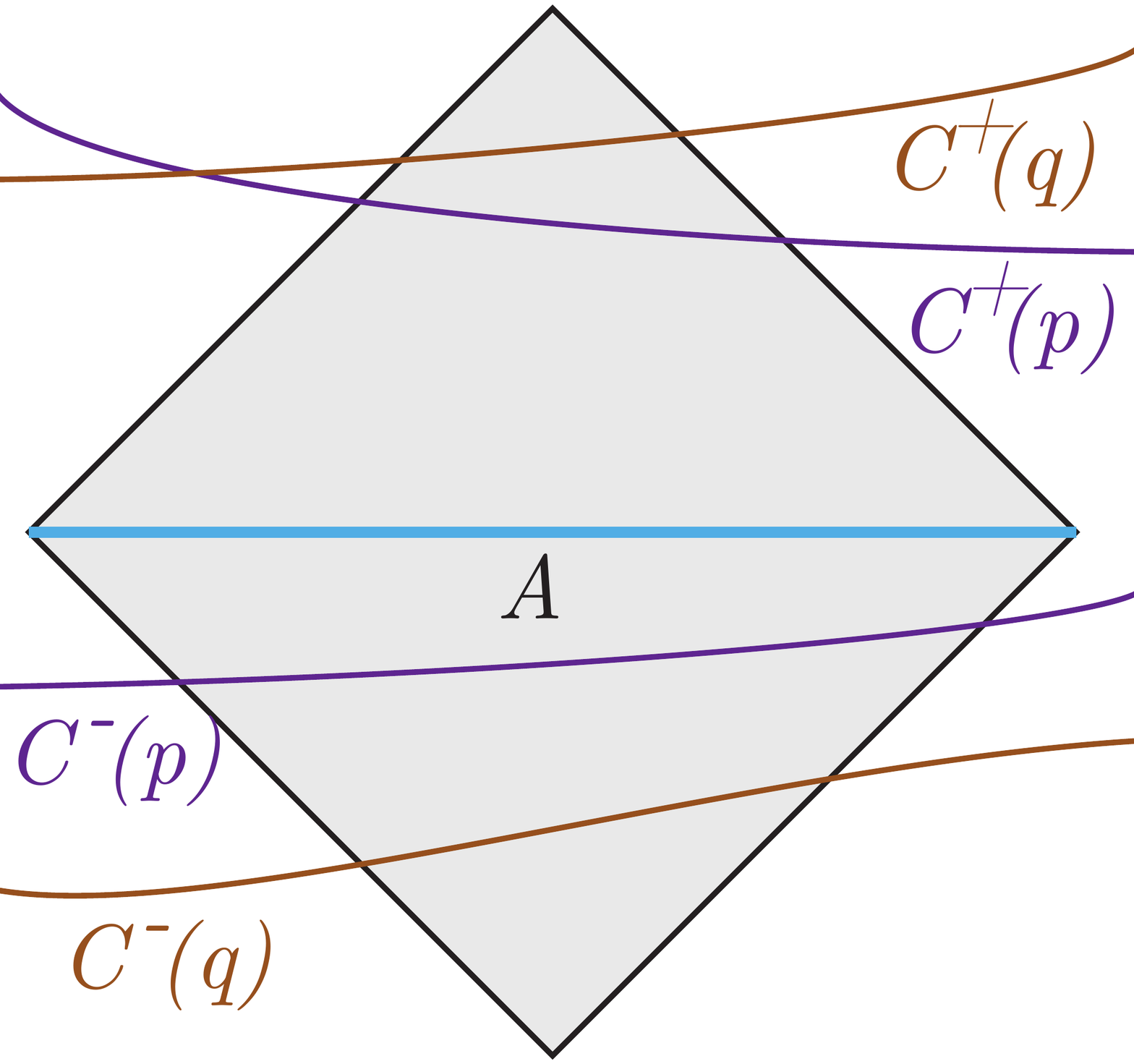}
\label{subfig:IntroRelativeAdS}
}
\caption{(a) A constant time slice of pure AdS. The region $A$ perceives $q$ as deeper in the bulk while the region $B$ perceives $p$ as deeper. (b) The lightcone cuts of $p$ (purple) are ``sandwiched'' by the lightcone cuts of $q$ (orange) on the domain of dependence of $A$ (gray): a larger subset of the domain of dependence of $A$ is spacelike to $q$ than to $p$, thus $q$ appears to be deeper in the bulk relative to $A$.}
\label{fig:IntroAdS}
\end{figure}

This intuition raises a conundrum in spacetimes where the entire bulk is in the causal wedge of the boundary (i.e. when there is no event horizon): for almost any pair of spacelike-separated bulk points, different boundary subregions provide different answers to the extent of causal contact they have with the points in question. This is illustrated in Fig.~\ref{subfig:IntroAdSSlice} for the case of pure AdS, where the point $p$ is perceived as deeper in the bulk than $q$ by the boundary subregion $A$, with the opposite result for the subregion $B$. 

\begin{figure}[t]
\centering
\subfigure[]{
\includegraphics[width=6cm]{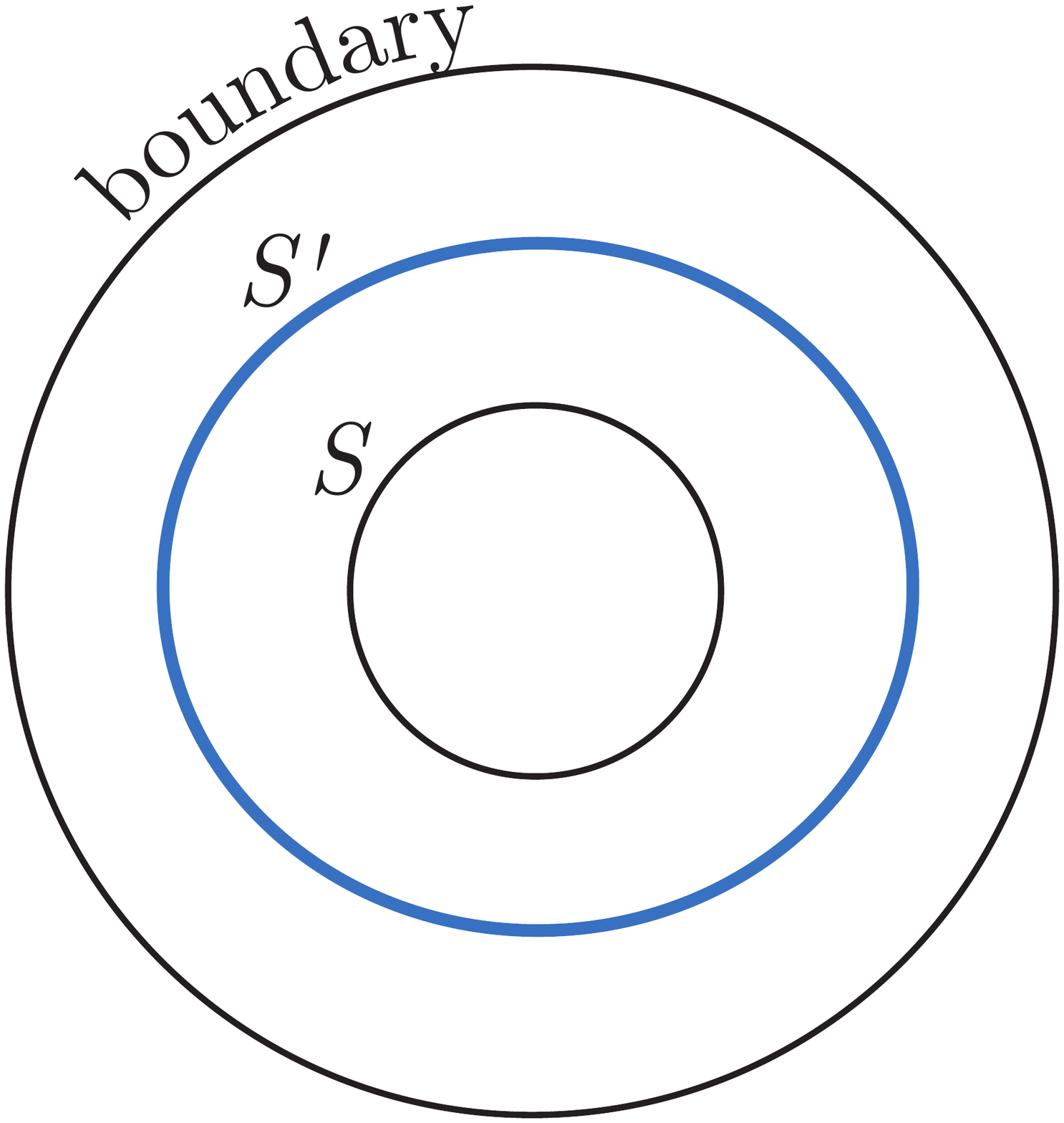}
\label{subfig:SurfaceDefSlice}
}
\hspace{1cm}
\subfigure[]{
\includegraphics[width=6cm]{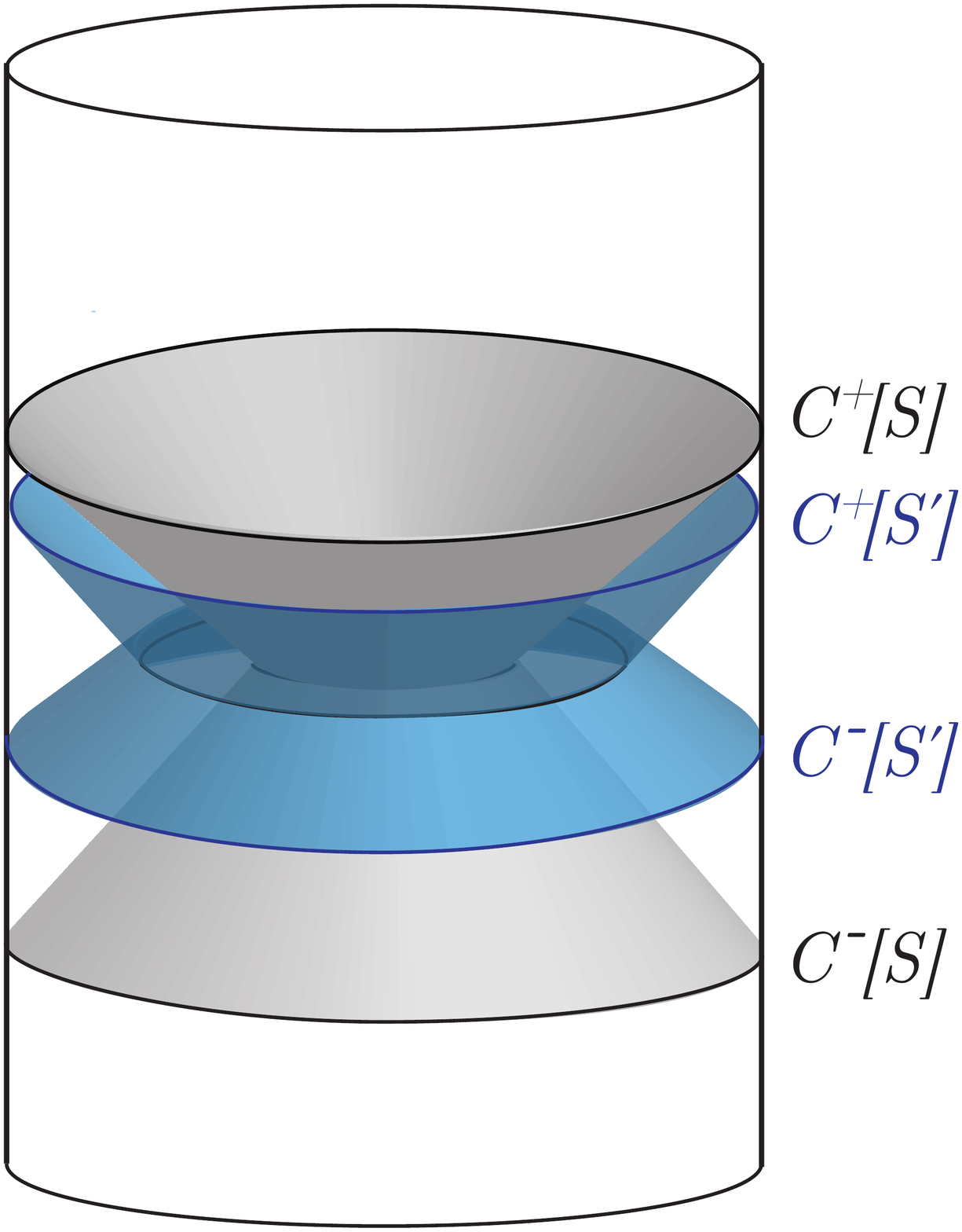}
\label{subfig:SurfaceDefFinal}
}
\caption{(a) A time slice of the bulk, where $S$ (black) and $S'$ (blue) are two spacelike surfaces on the same Cauchy slice. (b) The analogue of lightcone cuts for two surfaces $S$ (gray) and $S'$ (blue). The cuts of $S'$ are sandwiched by the cuts of $S$: a larger boundary proper time passes between $C^{\pm}[S]$ than between $C^{\pm}[S']$. $S$ is thus deeper in the bulk than $S'$.}
\label{fig:SurfaceDef}
\end{figure}

In such spacetimes, there is no absolute definition of comparative bulk depth for pairs of points: the only definition with physical content is one which qualifies depth relative to a fixed boundary subregion. Alternatively, in the absence of an event horizon it can be illuminating to consider the bulk depth of surfaces. For instance, we would like to consider a surface in pure AdS on some constant time slice and constant radius to be ``deeper in the bulk'' than a surface on the same time slice at larger radius. In this case, the intuition above works well: null geodesics fired from the ``deeper'' surface define a boundary slice to the future of those fired from the surface closer to the boundary. See Fig.~\ref{fig:SurfaceDef}.

\begin{figure}[t]
\centering
\includegraphics[width=8cm]{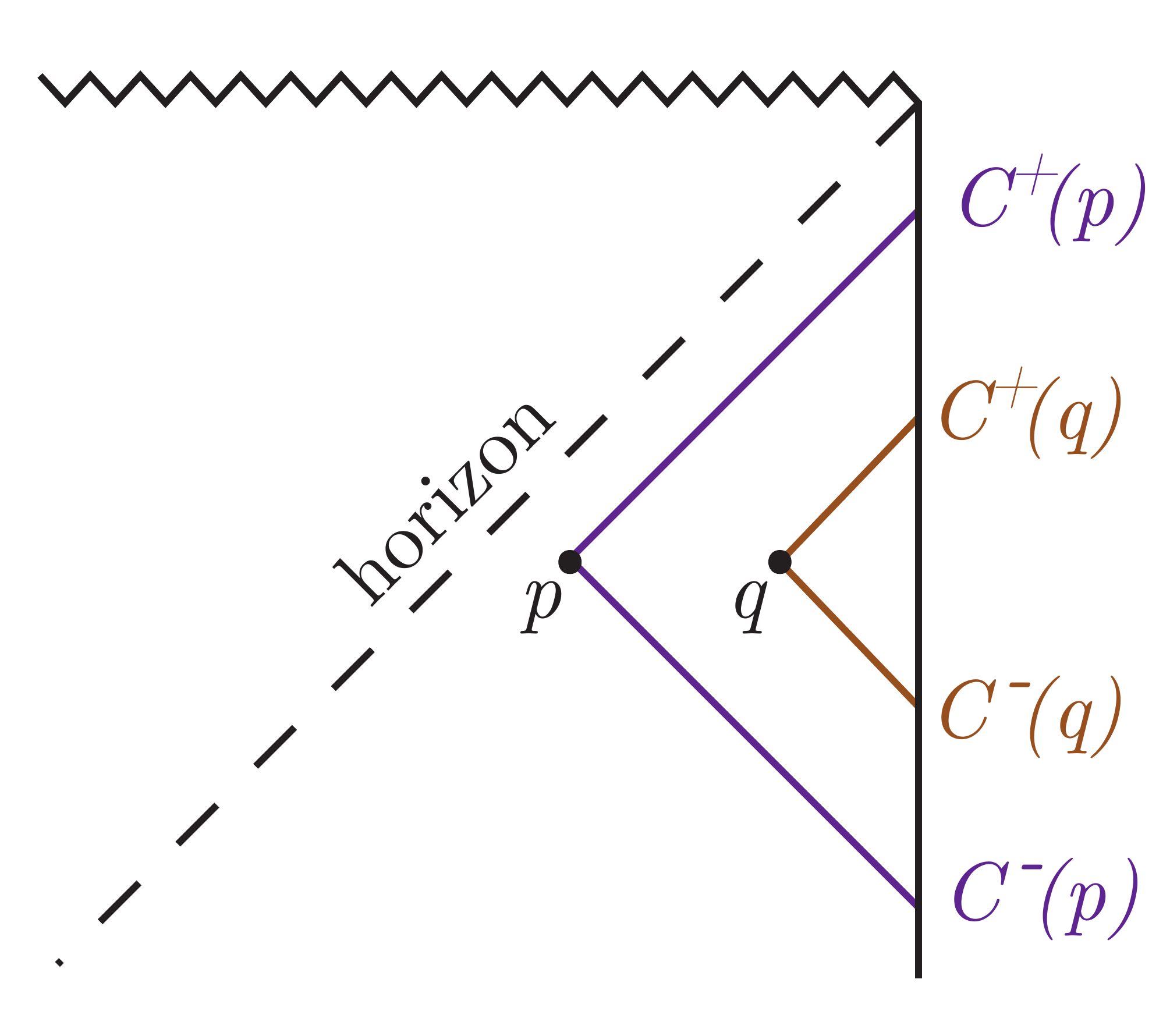}
\caption{In the presence of an event horizon, a point $p$ can be spacelike separated from a larger subset of the entire boundary than $q$. The result is a ``sandwich'' of the cuts of $q$ by the cuts of $p$. $p$ is deeper in the bulk relative to the \textit{entire} boundary.}
\label{fig:sandwichHorizon}
\end{figure}

When an event horizon exists, the intuition changes: it appears clear that points on the event horizon should be defined as ``deeper in the bulk'' than any spacelike-separated counterpart in the causal wedge of the boundary. In the presence of an event horizon, moreover, points that approach the event horizon are indeed spacelike-separated from a strictly larger subset of the entire boundary than points in the asymptotic region. See Fig.~\ref{fig:sandwichHorizon} for an illustration. 

We give a definition that allows for the comparison of bulk depth between any two bulk regions (points or surfaces) relative to a boundary region. The case where some points are deeper in the bulk than others relative to the entire boundary is simply a special case of our definition where the boundary region in question is the entire boundary. Our definition is simple: if, in some boundary causal diamond $D$, the subset of $D$ which is spacelike to some bulk region $S$ is properly contained in the subset of $D$ which is spacelike to some bulk region $S'$, then $S'$ is deeper in the bulk than $S$. This may be rephrased in terms of the lightcone cuts of points and their analogues for higher dimensional surfaces: $S'$ is deeper than $S$ relative to $D$ if the lightcone cuts of $S'$ ``sandwich'' the lightcone cuts of $S$ on $D$. Here by the lightcone cuts of a surface, we mean $\partial J^{\pm}[S]\cap \partial M$. See Fig.~\ref{fig:SurfaceDef} for an illustration. This definition is a \textit{partial} ordering of spacelike-separated points and surfaces by depth: certain bulk points will be located at the same depth.

%GETTING CUTS FROM CORRELATORS WILL NOT WORK EVERYWHERE IN AN ETERNAL BLACK HOLE DUE TO CLOSED NULL GEODESICS. THE DEFINITION WE GIVE IS STILL VALID, HOWEVER THE LIGHTCONE CUTS WILL NEED TO BE DETERMINED FROM OTHER OBSERVABLES. IN A COLLAPSING BLACK HOLE, A PRESCRIPTION EXISTS FOR GETTING THE CUTS (precite me and Gary). 

%
%Our definition is therefore twofold: it consists of a global criterion for qualifying when $p$ is absolutely deeper than $q$ (i.e. from the perspective of the entire boundary), and a local criterion for qualifying when $p$ is deeper than $q$ relative to a particular boundary subregion. The definition we propose is roughly the following: $p$ is (globally) deeper in the bulk than $q$ if the cuts of $p$ ``sandwich'' the cuts of $q$, i.e. $C^{+}(p)$ is to the future of $C^{+}(q)$, and $C^{-}(p)$ is to the past of $C^{-}(q)$. The point $p$ is deeper than $q$ relative to a boundary subregion ${\cal A}$ if the cuts of $p$ sandwich the cuts of $q$ on the domain of dependence of ${\cal A}$. See Fig.~\ref{fig:IntroAdS} for an illustration. 

Besides applicability to comparative depth perception of points and surfaces relative to a particular boundary subregion and to the entire boundary, a good definition of bulk depth must furthermore satisfy the following requirements: it must \textit{(i)} be covariant in a general bulk geometry,  \textit{(ii)} agree with known examples in which an obvious holographic coordinate exists and with general intuition that larger separation on the boundary corresponds to a deeper region in the bulk, and \textit{(iii)} have a well-understood field theory dual with a connection to energy scale. 

The above definition is clearly covariant and general. It turns out that motion along the holographic ``$r$'' coordinate in a static, spherical, hyperbolic, or planar AdS black holes agrees with our definition with motion deeper into the bulk: in such geometries, nested constant $r$ surfaces correspond to motion deeper in the bulk, in agreement with the UV/IR correspondence. 

We now turn to \textit{(iii)}: the location of lightcone cuts of points may be determined from the singularity structure of Lorentzian correlators, with some caveats. The procedure for obtaining the cuts from bulk-point singularities as outlined in~\cite{EngHor16a} fails to construct the cuts outside of the causal wedge, and also within the causal wedge for a region near the black hole event horizon. A  variant of the procedure exists for reconstruction up to the event horizon for black holes formed from collapse~\cite{EngHorTA}. We will proceed here under the assumption that we have obtained the lightcone cuts in some way or other even in eternal black hole geometries. Given the lightcone cuts of points in the causal wedge, there is an explicit procedure for obtaining to the bulk conformal metric. This allows us to construct the lightcone cuts of surfaces, which may also be found by taking the outer envelope of the lightcone cuts of their constituent points. 

Under assumption that we have been able to recover the lightcone cuts for the subset of the bulk of interest, the definition has a clear interpretation in the dual field theory. When a point $p$ is deeper than a point $q$, the correlators of interest are singular at longer time separations, which corresponds to lower energies; we recover a precise version of bulk point depth as a \textit{local} probe of dual infrared physics. A similar interpretation holds for bulk depth of surfaces.

Increasing time separation is perhaps a more unconventional manifestation of reaching infrared physics than increasing distance scales. It is natural to ask whether the definition above reproduces the standard intuition that larger spatial separations in the dual QFT correspond to increasing bulk depth. We find that this is indeed the case: a point $p$ is deeper in the bulk than a point $q$ if and only if every causal wedge containing $p$ also contains $q$. A similar statement holds for the relative bulk depth criterion.

The paper is structured as follows: Section~\ref{sec:rev} reviews the lightcone cut reconstruction; readers familiar with~\cite{EngHor16a} may wish to skip it. Section~\ref{sec:dep} formally introduces the definition of comparative bulk depth, as well as several useful constructs and a proposed measure of bulk depth; we prove that event horizons are sufficient for the global definition of comparative bulk depth to apply. Sec.~\ref{sec:AdS} gives an example of comparative bulk depth relative to a boundary subregion in pure AdS; Sec.~\ref{sec:BH} presents an argument that our definition of bulk depth of surfaces agrees with the holographic coordinate in AdS black holes with certain symmetries. Section~\ref{sec:causal} links our definition with depth in the bulk as measured by the causal wedges of boundary regions. We conclude in Section~\ref{sec:dis} with a discussion of possible applications for future directions and more speculative ideas.\\
%modified version of this theorem can be proved for the local bulk depth criterion.

%Our definition thus satisfies \textit{(i)} through  \textit{(iv)} above.

%In the language of lightcone cuts, $p$ is deeper than $q$  (with a restriction to the domain of dependence of a subregion for the local definition). See Fig. for an illustration. RELATIVE DEFINITION AND SURFACE DEFINITION (pastmost boundary of the union of lightcone cuts).

%relative to a boundary subregion $A$ if, in the (boundary) domain of dependence of $A$, the set of points causally accessible to $p$ is a proper subset of the set of points causally accessible to $q$. Globally, $p$ is deeper than $q$ if the subset of the boundary that is causally accessible to $p$ is properly contained in the subset causally accessible to $q$. 

%Gravitational dressing of points via a spacelike geodesic from the boundary has also been proposed as a definition of depth~\cite{}, but this approach raises other concerns: what is the QFT dual of the geodesic? How is the length of the geodesic regulated? At which boundary point should the geodesic be anchored? Rather than attempting to answer these questions, I will define a different way of qualifying bulk depth which circumvents them.

\noindent \textbf{Assumptions:} Thoughout this paper, we will assume that the bulk $(M,g)$ is a $(d+1)$-dimensional manifold, which is $C^{2}$, AdS hyperbolic\footnote{See~\cite{Wal12} for a definition.}, asymptotically AdS$_{d+1}$~\footnote{Most of the results below apply to asymptotically locally AdS geometries with the choice of a standard conformal frame on the boundary (as defined in~\cite{EngHor15}) and minor modifications, but for simplicity, we have focused exclusively on asymptotically AdS geometries.}, and obeys the Achronal Averaged Null Curvature Condition: $\int\limits_{\gamma} R_{ab}k^{a}k^{b}\geq 0$, where $R_{ab}$ is the Ricci tensor, and $k^{a}$ is the null generator of an achronal, complete geodesic $\gamma$. The symbol $\partial M$ is used to refer exclusively to a connected component of the conformal boundary. We do not assume the null generic condition. All results below apply in the large $N$, large $\lambda$ limit. All conventions are as in~\cite{Wald} unless otherwise stated.

\section{Review of Lightcone Cut Formalism}\label{sec:rev}
This section gives a review of the relevant aspects of the lightcone cut construction of~\cite{EngHor16a}. Those readers who are familiar with it may wish to skip to Sec.~\ref{sec:dep}. 

We begin by reminding the reader of some useful terminology: the causal future of a point $p$, denoted $J^{+}(p)$, is the collection of points that can be reached from $p$ by a timelike or null future-directed path. The causal past of $p$, $J^{-}(p)$ is simply the time reversed definition.

The future lightcone cut, or future cut for short, of a bulk point $p$, denoted $C^{+}(p)$,  is the set of earliest points on $\partial M$ (as measured by an observer on $\partial M$) that can be reached by null, future directed paths from $p$. The past lightcone cut of $p$, denoted $C^{-}(p)$, is defined similarly, in terms of null past-directed paths from $p$. Formally,

\begin{equation} C^{\pm}(p) \equiv\partial J^{\pm}(p)\cap \partial M.
\end{equation}

When a statement applies equally to past or future cuts, we will write $C(p)$.

The lightcone cuts of bulk points obey several useful properties, which we will use throughout this paper:
\begin{enumerate}
	\item $C^{\pm}(p)$ is a complete spatial slice of $\partial M$;
	\item $C^{\pm}(p)$ is a continuous set. $C^{\pm}(p)$ is furthermore $C^{1}$ everywhere except on at most a measure zero set~\cite{EngHorTA};
	\item $C^{\pm}(p)$ correspond to a unique bulk point $p$; $C^{\pm}(p)$ and $C^{\pm}(q)$ agree on an open set if and only if $C^{\pm}(p)=C^{\pm}(q)$, which is equivalent to $p=q$;
	\item Certain configurations of lightcone cuts correspond exclusively to spacelike-separated points. 
\end{enumerate}

\begin{figure}[t]
\centering
\subfigure[]{
\includegraphics[width=6cm]{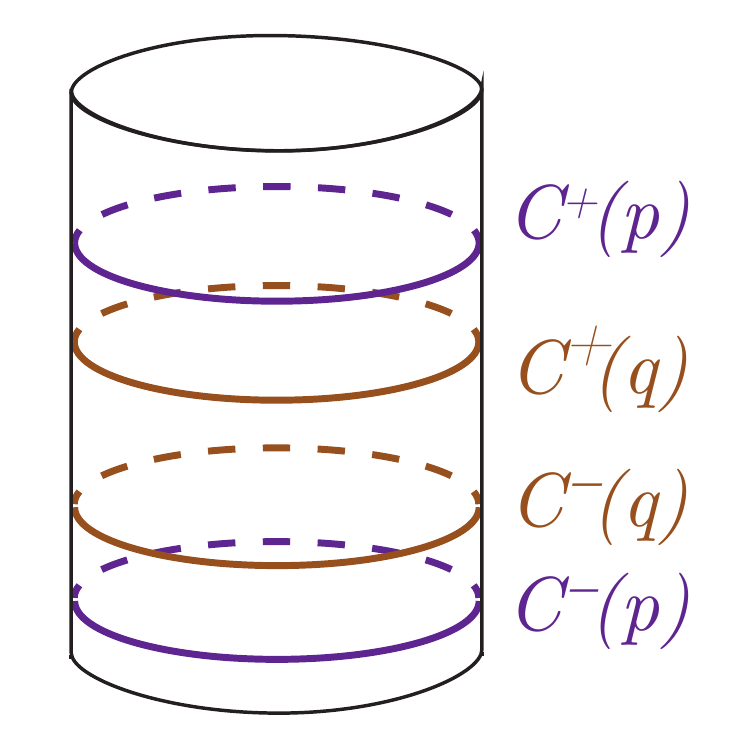}
\label{subfig:sandwichdef}
}
\hspace{1cm}
\subfigure[]{
\includegraphics[width=6cm]{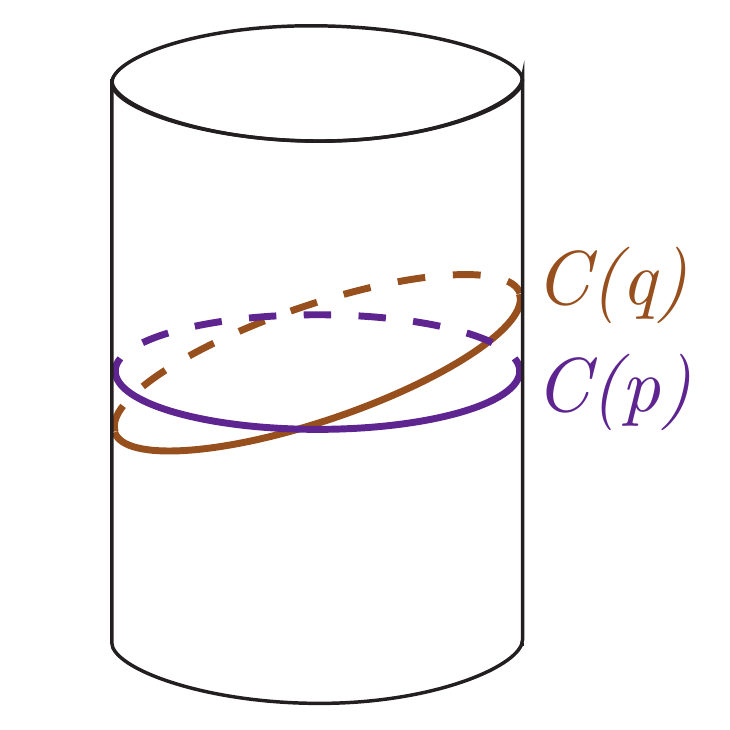}
\label{subfig:crossingdef}
}
\caption{Reproduced from~\cite{EngHor16a}. (a) The sandwich configuration. The lightcone cuts of $p$ (purple) are sandwiched by the lightcone cuts of $q$ (orange). (b) The crossing configuration. At least one of $C^{\pm}(p)$ intersects the corresponding cut of $q$.}
\label{fig:intersections}
\end{figure}

Particular use will be made of property (4). There are two specific configurations of the lightcone cuts of a pair of bulk points $\{p, q\}$, which must correspond to spacelike separated points:
\begin{itemize}
	\item ``Sandwich'': if $C^{+}(p)\subset I^{+}[C^{+}(q)]$ and $C^{-}(p)\subset I^{-}[C^{-}(q)]$, then $p$ and $q$ are spacelike-separated. The cuts of $p$ are said to \textit{sandwich} the cuts of $q$. See Fig.~\ref{subfig:sandwichdef}.
	\item ``Crossing'': if $C(p)$ intersects both $I^{+}[C(q)]$ and $I^{-}[C(q)]$, then $p$ and $q$ are spacelike-separated. The cuts of $p$ and $q$ are said to \textit{cross}. See Fig.~\ref{subfig:crossingdef}.
\end{itemize}

The sandwich configuration is of primary interest: if the cuts of $p$ sandwich the cuts of $q$, then the subset of $\partial M$ which is spacelike to $q$ is properly contained in the subset of $\partial M$ which is spacelike to $p$. That is, any boundary curve will cease to have causal contact with $p$ while still in causal contact with $q$, and will regain causal contact with $q$ before regaining causal contact with $p$.  The sandwich configuration will be used in the next section to qualify when one bulk point is deeper than another relative to the entire boundary.

Ref. \cite{EngHor16a} gave an explicit procedure for reconstructing the bulk conformal metric from lightcone cuts\footnote{The bulk conformal metric is the equivalence class of metrics in the bulk, which are all related by an overall rescaling: $\Omega^{2}g\sim g$.}. If we are given the cuts for some subset of the bulk, we may therefore safely assume that the bulk conformal metric is known data within that subset. This immediately grants access to causal separation between points and surfaces; we know when $p$ and $q$ are spacelike-separated, and we can construct the entirety of their lightcone within the causal wedge --- provided that we have been able to determine the location of the lightcone cuts.

Next, we review a procedure of obtaining the lightcone cuts from the dual field theory suggested by~\cite{EngHor16a}. It is possible other, more general ways of determining the cuts exist. This particular procedure makes use of the singularity structure of Lorentzian correlators in the dual field theory.

A time-ordered Lorentzian $(d+3)$-point correlator\footnote{In~\cite{MalSimZhi}, $(d+2)$-point correlators were considered; the additional point is required for the construction of~\cite{EngHor16a}. Since in this work we will be using the latter, we focus on $(d+3)$-point correlators.} $\langle \mathcal{O}(x_{1})\cdots \mathcal{O}(x_{d+3})\rangle$ of some operator $\mathcal{O}$ is singular when the points $\{x_{1},\cdots x_{d+3}\}$ are null-separated from a common vertex and energy-momentum is conserved at the vertex. In a holographic QFT dual to a semiclassical geometry, this vertex may lie in the bulk. A singularity in the $(d+3)$-point correlator which is not sourced by boundary vertex thus identifies a bulk point; such singularities are called \textit{bulk-point singularities}~\cite{MalSimZhi}. 

A bulk point $p$ can then be identified with two time-separated spatial slices on the boundary: the two spatial boundary slices with the smallest time separation at which the $(d+3)$ correlators are singular due to a bulk vertex. These slices are precisely the intersection $C^{\pm}(p)=\partial J^{\pm}(p)\cap \partial M$. 

The lightcone cuts determine the bulk geometry only up to an overall function\footnote{Extensions which include bulk spacetimes where the Einstein tensor is traceless will appear in~\cite{EngHorTA}.}. If we wish to adhere purely to this formalism without assuming that we know the bulk geometry, this may \textit{prima facie} appear to be an insurmountable hurdle. Fortunately, our goal is determine whether $p$ is deeper in the bulk than $q$; we do not attempt to give a measure of precisely how deep $p$ is, although a potential approach to this problem is discussed in Section~\ref{sec:dep}. Such an approach works under the assumption that the bulk conformal factor has been determined in some way. For a comparison of relative depth between two points, the conformal factor is unnecessary.

\section{Bulk Depth from Lightcone Cuts}\label{sec:dep}

We will now use the lightcone cut formalism described in the previous section to precisely define when one bulk point is deeper than another. Before we proceed, let us first remind the reader of a few standard definitions.
\begin{itemize}
	\item Let ${\cal A}$ be a $(d-1)$-dimensional spacelike boundary region. The domain of dependence of ${\cal A}$ on the boundary is defined $D_{\partial}[{\cal A}]=D_{\partial}^{+}[{\cal A}]\cup D_{\partial}^{-}[{\cal A}]$, where $D_{\partial}^{+}[{\cal A}]$ ($D_{\partial}^{-}[{\cal A}]$) is the set of all points $p$ such that every boundary-contained past-directed (future-directed) causal curve through $p$ passes through ${\cal A}$. 
	\item The causal wedge of ${\cal A}$, denoted $C_{W}[{\cal A}]$, is defined as the set of bulk points which can send both past- and future-directed causal curves to $D_{\partial}[{\cal A}]$: $C_{W}[{\cal A}]\equiv J^{-}[D_{\partial}[{\cal A}]]\cap J^{+}[D_{\partial}[{\cal A}]]$~\cite{BouLei12, HubRan12}\footnote{The causal wedge is sometimes defined in the literature in terms of the chronological past and future $I^{\pm}[D_{\partial}[{\cal A}]]$. We find it more convenient here to use the causal past and future $J^{\pm}[D_{\partial}[{\cal A}]]$. All results quoted in this text about the causal wedge apply for either definition under our list of assumptions.}.
	\item The causal surface of ${\cal A}$, denote $C[{\cal A}]$ is defined as $C[{\cal A}]\equiv \partial J^{-}[D_{\partial}[{\cal A}]]\cap \partial J^{+}[D_{\partial}[{\cal A}]]$ \cite{HubRan12}.
\end{itemize}

We look for a definition of bulk depth that captures the idea that observers on the boundary spend a longer proper time out of causal contact with points that are deeper in the bulk. As noted above, such a definition should be adaptable to both global depth, where any boundary observer must spend longer out of causal contact with a point, and to relative depth, where this holds only for boundary observers in a certain causal diamond. 

To qualify bulk depth between surfaces, we extend the terminology of lightcone cuts to extended objects. Let $S$ be a spacelike bulk surface. The union of the future (past) lightcone cuts of all points in $S$ defines a $d$-dimensional boundary region. The outer envelope of that region, i.e. the past (future) boundary of that region, is  the future (past) lightcone cut of $S$:
\begin{align} &C^{+}[S]=\partial J^{+}[S]\cap \partial M\\
& C^{-}[S]= \partial J^{-}[S]\cap \partial M.
\end{align}
The definition of the sandwich and crossing configuration for the lightcone cuts of surfaces is identical to that for the lightcone cuts of points.

We are now ready to state the definition of comparative bulk depth between two bulk surfaces:\\

\noindent \textit{Comparative Bulk Depth:} Let ${\cal A}$ be a $(d-1)$-dimensional spacelike boundary subregion (where ${\cal A}$ is not a complete spacelike slice of $\partial M$). The region $S$ is said to be \textit{deeper in the bulk relative to ${\cal A}$} than the region $S'$ if on $D_{\partial}[{\cal A}]$, the cuts $C^{\pm}[S]$ sandwich $C^{\pm}[S']$. More precisely, if the both of the following conditions are satisfied:
\begin{align} & C^{+}[S]\cap D_{\partial}[{\cal A}] \subset I^{+}[C^{+}[S']]\cap D_{\partial}[{\cal A}] \\
 & C^{-}[S]\cap D_{\partial}[{\cal A}] \subset I^{-}[C^{-}[S']]\cap D_{\partial}[{\cal A}] ,
\end{align}
with the additional provision that, if the cuts of $S'$ intersect $D_{\partial}[{\cal A}]$ and $J^{\pm}[S]\cap D_{\partial}[{\cal A}]= \varnothing$ (so $D_{\partial}[{\cal A}]$ is spacelike to $S$ everywhere) then $S$ is deeper than $S'$ relative to ${\cal A}$.

This is illustrated in Fig.~\ref{subfig:IntroRelativeAdS}. Note that $S, S'$ can be spacelike surfaces of any dimension; in particular, $S$ and $S'$ can be points.

To determine whether $S$ is deeper than $S'$ relative to a boundary region ${\cal A}$, we must therefore construct the boundary domain of dependence of ${\cal A}$, and find the location of the lightcone cuts of $p$ and $q$ on this domain of dependence.

\begin{figure}[t]
\centering
\includegraphics[width=6cm]{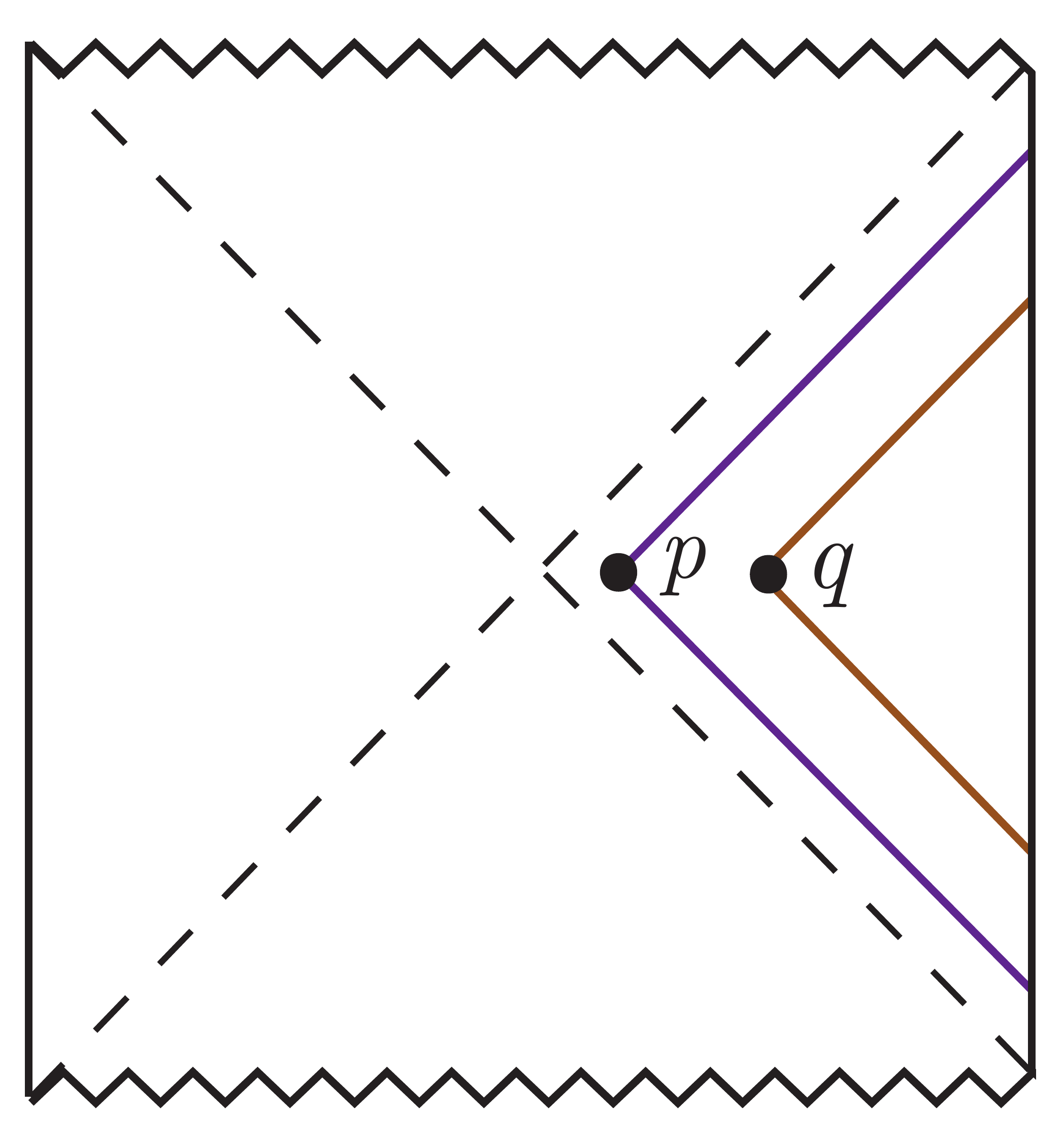}
\caption{In Schwarzschild-AdS, points near the bifurcation surface are deeper in the bulk than points in the asymptotic region. More precisely, points at smaller radial coordinate $r$ (outside of the horizon) are deeper in the bulk. Formally, the lightcone cuts of points on the bifurcation surface are at past and future infinity on the boundary in the static cylinder frame.}
\label{fig:Schwarzschild}
\end{figure}

The above definition captures precisely the notion that a larger component of the boundary or boundary subregion is spacelike to a point (or surface) which is deeper in the bulk. As illustrated in Fig.~\ref{fig:Schwarzschild}, points on the bifurcation surface of Schwarzschild-AdS are acausal to all boundary points at finite boundary time, marking the bifurcation surface as the deepest surface in the causal wedge of $\partial M$.\\

It is instructive to restate the definition in the special case when we are interested in \textit{global depth comparison}: determining when $S$ is deeper than $S'$ relative to the entire boundary. We further break the definition down into points and surfaces:\\

\noindent \textit{Global Depth of Points:} Let $p$ and $q$ be two spacelike-separated bulk points in the causal wedge of the entire boundary. The point $p$ is said to be \textit{deeper in the bulk} than the point $q$ if the cuts of $p$ sandwich the cuts of $q$, as defined in Sec.~\ref{sec:rev}. See Fig.~\ref{subfig:sandwichdef}. \\

\noindent \textit{Global Depth of Surfaces:} Let $S$, $S'$ be two bulk surfaces on the same Cauchy slice of $M$. $S$ is deeper in the bulk than $S'$ if $C^{+}[S]$ is in the future of $C^{+}[S']$ and $C^{-}[S]$ is in the past of $C^{-}[S']\cap \partial M$. See Fig.~\ref{fig:SurfaceDef} for an illustration.\\

Recast in this form, the definition clearly has the desired consequence that, if $S\subset \text{Int}[S']$, (where the interior here is defined on some Cauchy surface), then $S$ by this definition lies deeper in the bulk than $S'$, when both surfaces live entirely within the causal wedge of $\partial M$. \\

\noindent \textbf{Existence of Sandwiches:} The reader may at this point protest that the sandwich cut configuration is one of several possible cut configurations for spacelike-separated bulk points. When is there a guarantee that the sandwich configuration exists? We have claimed that the sandwich is a natural configuration in black hole geometries. Indeed, the existence of an event horizon is sufficient for the sandwich configuration to exist.\\

\noindent \textit{Theorem:} There exist pairs of bulk points $\{p,q \}$ whose lightcone cuts form a sandwich whenever $C_{W}[\partial M]\subsetneq M$. 

\begin{proof} To show this, it is sufficient to construct a pair of points whose lightcone cuts form a sandwich in a spacetime with an event horizon, i.e. when $C_{W}[\partial M]\subsetneq M$. 

Assume without loss of generality that the connected component ${\cal H}^{+}=\partial J^{-}[\partial M]$ is nonempty, so there is a future event horizon; the arguments below apply equally well to past horizons under time reversal. Generators of $\partial J^{-}[\partial M]$ do not intersect $\partial M$ once they have entered the bulk, by the Gao-Wald theorem~\cite{GaoWal00} and because  $\partial J^{-}[\partial M]$ is achronal, and as assumed above, $\partial M$ is a single connected component of the boundary. 

Let $p$ be a point on ${\cal H}^{+}$, so that $C^{+}(p)$ is formally at future infinity in the boundary spacetime, while $C^{-}(p)$ is generally at some finite time, see Fig.~\ref{fig:proof} (the arguments below apply equally well if $p$ lies on the bifurcation surface, and $ C^{-}(p)$ is at infinite past time in the boundary). Let $t_{0}$ be the largest value of the boundary time coordinate (in the Einstein Static Universe frame) on $C^{-}(p)$, and consider firing a future-directed null geodesic $\gamma$ into the bulk at finite boundary time $t>t_{0}+ \pi$. Let $q$ be any point on $\gamma$ in the boundary causal wedge $C_{W}[\partial M]$. By construction, there are future-directed paths from $q$ to $\partial J^{+}(p)$ and past-directed paths from $q$ to $\partial J^{-}(p)$, which immediately implies that $q$ and $p$ are spacelike separated. 

If $\gamma$ is achronal from $q$ to $\partial M$, then $\gamma$ is a generator of $\partial J^{-}(q)$; if $\gamma$ is chronal from $\partial M$ to $q$, then $\gamma\cap \partial M$ lies in the past of $C^{-}(q)$. See Fig.~\ref{fig:proof}.
\begin{figure}[t]
\centering
\includegraphics[width=7cm]{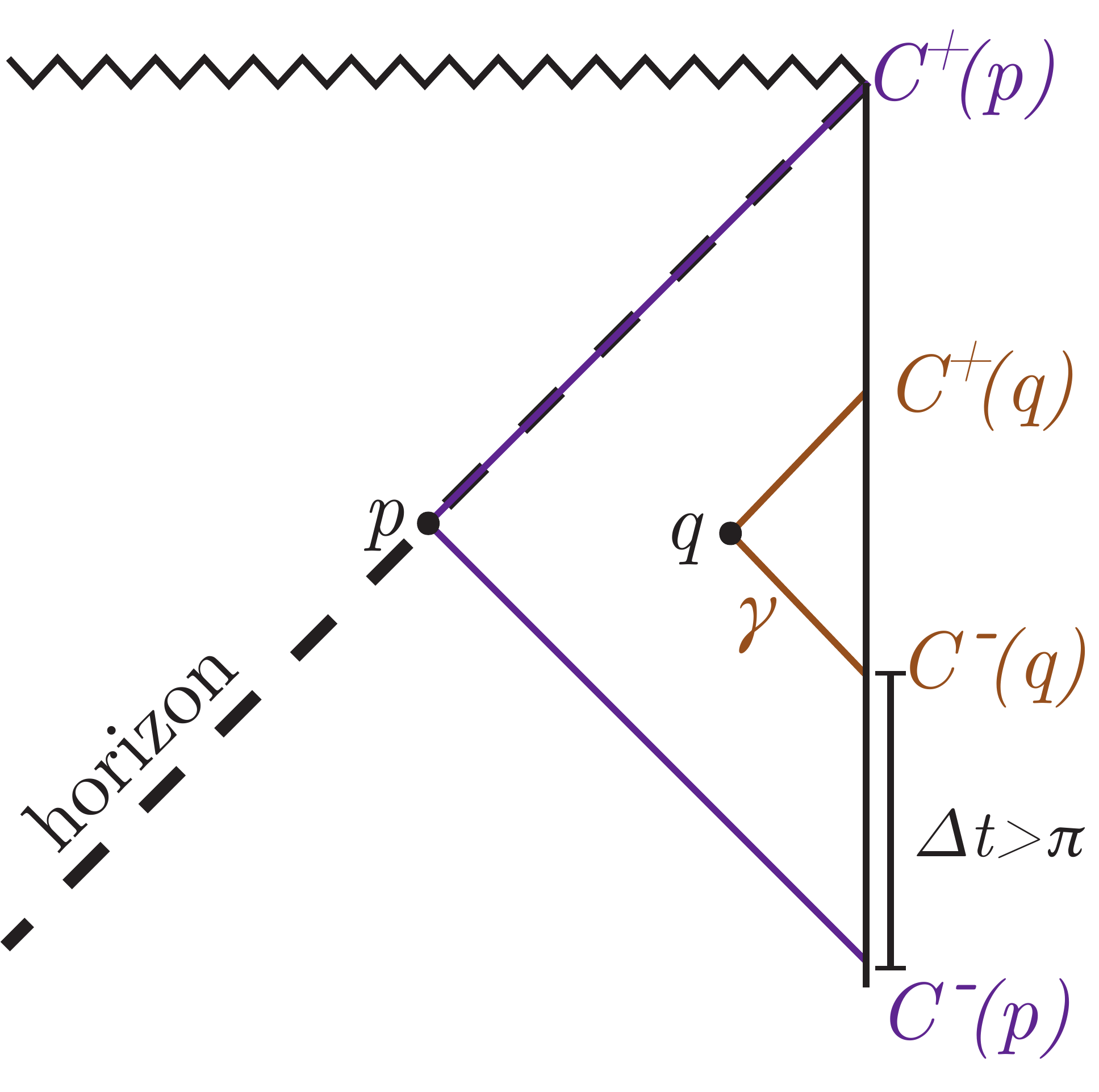}
\caption{Proof of sufficiency of the event horizon for the existence of sandwich configurations. If $\gamma$ is chronal, $C^{-}(q)$ lies farther in the future than illustrated here.}
\label{fig:proof}
\end{figure} 
Either way, $C^{-}(q)$ has points at $t>t_{0}+\pi$; because $\Delta t=\pi$ is the lightcrossing time on the boundary, this means that every point on $C^{-}(p)$ is timelike-separated from $C^{-}(q)$: $C^{-}(p)\subset I^{-}[C^{-}(q)]$. Since $q$ is by construction spacelike-separated from $p$ and within the interior of the causal wedge, $C^{+}(q)$ lies at finite boundary coordinate time. Formally, the cuts of $q$ are sandwiched between the cuts of  $p$. To get a sandwich for two points, both with cuts at finite boundary time, we can also deform $p$ into $C_{W}[\partial M]$. Because $I^{\pm}$ is an open set, sufficiently small deformations will leave the sandwich structure intact. 

\end{proof}

Intuitively, it seems likely that the existence of an event horizon, or at the very least  null geodesics that do not reach $\partial M$, should also be a necessary condition for the sandwich configuration: this configuration captures the notion of progressively fewer lightcone generators reaching the boundary when the point is moved in a spacelike direction. This intuition is discussed further in Sec.~\ref{sec:causal}. \\

\noindent \textbf{Efficient Curves:} The discussion above has focused on  determining whether and when one of two bulk points can be qualified as living deeper in the bulk. We have thus far ignored the question of the measurement of such depth. 

\begin{figure}[t]
\centering
\includegraphics[width=6cm]{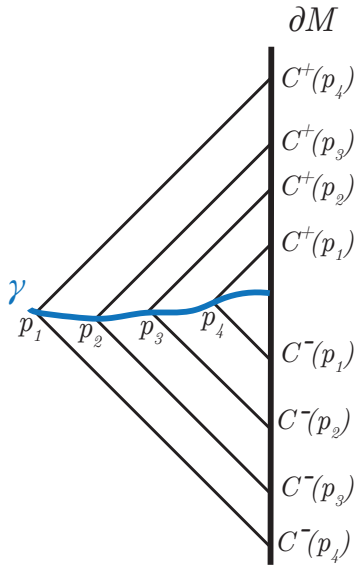}
\caption{An efficient curve $\gamma$ (blue) moves continuously through points with bulk depth increasing monotonically along $\gamma$.}
\label{fig:efficient}
\end{figure}

Is there a natural object that measures the depth between two points? Certainly the proper time elapsed on the boundary between cuts is a possible candidate, however it changes with different choices of a boundary conformal frame. We propose instead to use distinguished bulk curves we term \textit{efficient curves}. The shortest such efficient curve between two points, when it exists, provides a natural measure of the comparative depth between a point $p\in S$ and $q\in S'$. \\

\noindent \textit{Efficient Curve}: Let $p$ and $q$ be two spacelike-separated regions or points in $M$, and let $p$ be deeper than $q$ (globally or relative to a subregion ${\cal A}$). If there exists a spacelike bulk curve $\gamma(s)$ between $p$ and $q$, $\gamma(1)=p$ and $\gamma(0)=q$, such that any point $\gamma(s_{1})$ is deeper (globally or relative to ${\cal A}$) than $\gamma(s_{2})$ when $1>s_{1}>s_{2}>0$, then we say such a curve $\gamma$ is an \textit{efficient curve}. See Fig.~\ref{fig:efficient} for an illustration. We define the relative depth between $p$ and $q$ as the length of the minimal efficient curve between them.\\

The intuition behind the definition of the efficient curve is that \textit{locally} it is the most efficient way to move deeper into the bulk, since it moves deeper monotonically. It is possible that there is a shorter distance curve between $p$ and $q$ than the minimal length efficient curve between them, but it will not move into the bulk in a locally efficient way. Note that the existence of an efficient curve between two surfaces $S$ and $S'$, where $S$ is deeper than $S'$, implies that there is a point on $S$ that is deeper than (or at the same depth as) any other point on $S$.

It is not clear that efficient curves always exist; it will be shown below, however, that radial geodesics from the boundary are efficient curves in static AdS black holes with isotropic spatial slices. Moreover, radial geodesics in black holes with such symmetries are, in fact, efficient curves. Generically, efficient curves need not be geodesics; an efficient curve is defined using only lightcone cuts, which are conformal invariants of the bulk geometry. Spacelike geodesics are not conformal invariants, so while an efficient curve may be a geodesic with one choice of bulk conformal factor, it will generically not be a geodesic with a different choice.  

Finally, we note a quick caveat: in order to measure the lengths of efficient curves, we must have access to the full bulk geometry in the causal wedge. That is, both the conformal metric and the conformal factor must be known. Knowledge of the locations of lightcone cuts is not sufficient except in certain special cases. Assuming that the full bulk metric is known, we are free to measure lengths along curves in the bulk.

\subsection{Example: Pure AdS}\label{sec:AdS}

For simplicity, we consider AdS$_{3}$, although the calculation below can easily be adapted to higher dimensions. It is useful to work in global coordinates:
\begin{equation} ds^{2}= -(1+r^{2})dt^{2} + \frac{dr^{2}}{1+r^{2}} +r^{2}d\theta^{2},
\end{equation}
\noindent where the AdS length scale has been set to 1. We fix the boundary conformal frame to the Einstein Static Universe:
\begin{equation} ds^{2}_{\partial} = -dt_{\partial}^{2}+d\theta_{\partial}^{2},
\end{equation}
\noindent where the subscripts on the coordinates serve as a reminder that these are boundary coordinates.  

The lightcone cuts of pure AdS (in any dimension) can be obtained simply from symmetry considerations. In terms of boundary coordinates, the lightcone cuts of a bulk point at $t=t_{0}$, $r=r_{0}$, $\theta=\theta_{0}$ are~\cite{EngHor16a}:
\begin{equation} \tan(t_{\partial} -t_{0})= \pm \frac{1}{r_{0}\cos(\theta_{\partial}-\theta_{0})}\left [1+r_{0}^{2}\sin^{2}(\theta_{\partial}-\theta_{0}) \right]^{1/2}.
\end{equation}

\begin{figure}[t]
\centering
\includegraphics[width=7cm]{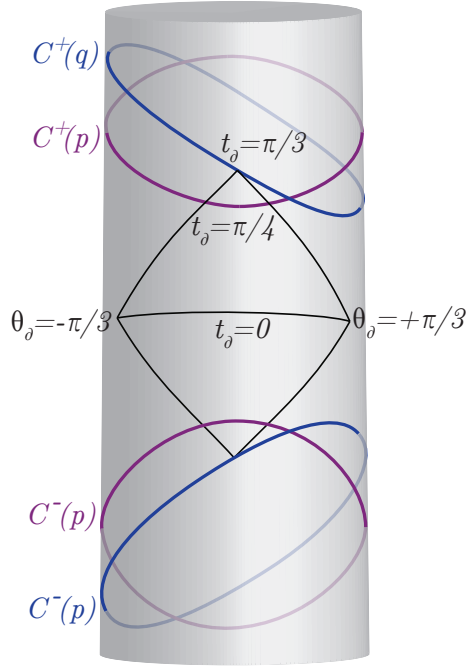}
\caption{The lightcone cuts of $(t_{0}, \theta_{0},r_{0})=(0,0,1)$ and $(0, \pi/4,1)$ in pure AdS$_{3}$, drawn with the domain of dependence of the boundary region $t_{\partial}=0$, $\theta\in [-\pi/3, \pi/3]$. Relative to this region, $C^{\pm}(q)$ sandwich $C^{\pm}(p)$: $q$ is deeper in the bulk.}
\label{fig:egAdS}
\end{figure} 

\noindent Consider two bulk points $p$ and $q$ at $t_{0}=0$ and the same radius $r_{0}=1$, and at different angular positions: $\theta_{0}=0$ and $\theta_{0}=\pi/4$, respectively. The lightcone cuts of $p$ and $q$ are illustrated in Fig.~\ref{fig:egAdS}. As is clear from the figure, the lightcone cuts of $p$ and $q$ cross. In fact, the lightcone cuts of any two spacelike-separated points in pure AdS must cross~\cite{EngHor16a}, in agreement with the idea that there is no notion of absolute depth of points in pure AdS (and as noted above more generally, we might expect the same in any causally trivial spacetime).

It is simple to identify boundary subregions that perceive cuts of $q$ as sandwiching the cuts of $p$ or vice versa. To wit, we have illustrated the boundary domain of dependence of a region that perceives $q$ ($\theta_{0}=\pi/4$) as deeper in the bulk than $p$ ($\theta_{0}=0$). For a region ${\cal A}$ at $t_{\partial}=0$, $\theta_{\partial}\in[-a,a]$, $D_{\partial}[{\cal A}]$ intersects $C^{\pm}(p)$ when $a>\pi/4$; between $a=\pi/4$ and $a=\pi/3$, $D_{\partial}[{\cal A}]$ intersects $C^{\pm}(p)$ and not $C^{\pm}(q)$. For $a\in (\pi/3,  \pi/8+\tan^{-1}(\sqrt{7-4\sqrt{2}}))$, the cuts of $q$ sandwich the cuts of $p$. This is illustrated in Fig.~\ref{fig:egAdS}.  At $a=\pi/8+\tan^{-1}(\sqrt{7-4\sqrt{2}})$, $D_{\partial}[{\cal A}]$ includes an intersection between the cuts of $p$ and the cuts of $q$: at that point the region ${\cal A}$ perceives neither $q$ nor $p$ as deeper in the bulk. We can similarly obtain a region that perceives $p$ as deeper in the bulk than $q$ by rotating ${\cal A}$ above on the boundary sphere.

The depth of surfaces, however, is another matter. In order to make contact with the UV/IR correspondence as it is usually formulated, a constant $r$, constant $t$ surface should be deeper in the bulk than a surface at the same time slice and a larger value of $r$; this is illustrated in Fig.~\ref{fig:SurfaceDef}. It is easy to see that this is the case in pure AdS. Let $S(r_{0})$ be a surface at constant $r=r_{0}$ and constant time. Since radial, future-directed geodesics in AdS are the future-directed geodesics to reach the boundary at the smallest coordinate time $t_{\partial}$, the future lightcone cut of $S$ is generated by radial geodesics. The elapsed coordinate time along such geodesics is:
\begin{equation}
\Delta t =\int\limits_{r_{0}}^{\infty} \frac{dr}{r^{2}+1} = \frac{\pi}{2} -\tan^{-1}(r_{0}).
\end{equation}
\noindent The lightcone cuts of $S$ are therefore located at $t_{\partial}=t_{0}\pm (\pi/2 -\tan^{-1}(r_{0}))$. Because $\tan^{-1}(r_{0})$ increases as $r_{0}$ increases, reaching $\pi/2$ as $r_{0}$ reaches infinity, the cuts $C^{\pm}(r_{0})$ are sandwiched between the cuts of $C^{\pm}(r_{1})$ whenever $r_{1}<r_{0}$. 

\subsection{Example: Black Holes with Symmetry}\label{sec:BH}

Most of the intuition concerning bulk depth is a result of work in spacetimes with well-defined holographic coordinate. We would like to check that our covariant definition above agrees with the idea of bulk depth in such spacetimes. 

In any static, asymptotically AdS black hole spacetime whose preferred spatial slices are isotropic, the metric outside of an event horizon can be written as follows: 

\begin{equation}
ds^{2} =-f(r)dt^{2} +\frac{dr^{2}}{f(r)}+r^{2}d\Sigma^{2},
\end{equation}
\noindent where $d\Sigma^{2}$ is the line element of a $(d-1)$-dimensional symmetric space (e.g. sphere, hyperboloid, plane), and in particular it is independent of $r$; $f(r)$ vanishes at the event horizon. We have set the AdS radius to 1. In such spacetimes, it is natural to think of the $r$-direction as the ``holographic direction''. We would like our definition to qualify surfaces at larger values of $r$ as deeper in the bulk than surfaces at smaller values of $r$.
%
%Null geodesics in such spacetimes obey the following equation (within the causal wedge):
%\begin{equation}
%t'(r)^{2} =\frac{1}{f(r)^{2}} + \frac{r^{2}}{f(r)}\sigma'(r)^{2},
%\end{equation}
%\noindent where some abuse of notation has hopefully not obfuscated the meaning of $\sigma'(r)$. 

Let $S$ be a spacelike surface at $t=0$ and $r=r_{0}$. Because radial geodesics are the earliest to reach the boundary, the lightcone cut of $S$ is generated by radial geodesics. If a radial geodesic starting at $r=r_{0}$ takes time $\Delta t_{1}$ to reach the boundary, then a radial geodesic starting at $r<r_{0}$ takes a time $\Delta t_{2}>\Delta t_{1}$ to reach the boundary. This follows from the isotropy of the spatial slices: starting at $r>r_{0}$, we may simply follow the radial null geodesic to $r=r_{0}$; let the time elapsed to reach $r_{0}$ be $\Delta t'$. From $r_{0}$, it takes the geodesic $\Delta t_{1}$ to reach the boundary; it therefore immediately follows that $\Delta t_{2}=\Delta t'+\Delta t_{1}>\Delta t_{1}$. Time reversal symmetry allows us to apply the same argument to past and future cuts.

We therefore find that surfaces at constant values of $r$ must correspond to cuts in a sandwich configuration. This geometry thus admits efficient curves at constant $\theta$ and constant $t$; these are precisely radial spacelike geodesics at constant $t$.

%
%The time elapsed along these is given by: 
%\begin{equation}
%\Delta t(r_{0}) = \int\limits_{r_{0}}^{\infty} \frac{dr}{f(r)}.
%\end{equation}

%
%\noindent It is clear that the integral grows monotonically as we increase the interval of integration (i.e. when $r_{0}$ is decreased). By the time-reversal symmetry of these geometries, this immediately implies that surfaces at constant values of $r$ must correspond to cuts in a sandwich configuration. This geometry therefore admits efficient curves at constant $\theta$ and constant $t$; these are precisely radial spacelike geodesics at constant $t$. 

\section{Relation to the Causal Wedge}\label{sec:causal}
 
One approach towards probing deeper into the bulk is via the use of nonlocal boundary observables at progressively larger spatial separation. A particularly natural object to consider in light of the importance of causality to our definition is the causal holographic information of a boundary region ${\cal A}$. This object is defined as the area of the causal surface of ${\cal A}$~\cite{HubRan12}. It was shown in~\cite{HubRanTon} (see also~\cite{Rib05, Rib07} for earlier work) on the subject that the causal wedges of nested boundary regions are themselves nested, so that increasing the size of the boundary region results in correspondingly larger causal wedges. Any reasonable definition of bulk depth should therefore consistently qualify points in the causal wedge of a region ${\cal A}$ as deeper in the bulk than those in the causal wedge of a region ${\cal A'}$ whenever ${\cal A'}$ is a proper subset of ${\cal A}$.

As expected, our definition above is compatible with the above requirement: a point $p$ is deeper in the bulk than a point $q$ by our definition if and only if every causal wedge containing $p$ also contains $q$. A similar result can be shown for relative depth. These results rely on two lemmata:\\

\noindent \textit{Lemma 1:} If ${\cal A}$ and ${\cal B}$ are spacelike-separated, then $C_{W}[{\cal A}]$ and $C_{W}[{\cal B}]$ are spacelike separated (with no overlap). \\

\noindent This was proved in~\cite{HubRanTon}. In particular, it implies that if ${\cal A}=\bigcup\limits_{i}{\cal A}_{i}$ is the disjoint union of connected, closed, spacelike-separated, spacelike components ${\cal A}_{i}$, then $p\in C_{W}[{\cal A}]$ implies that $p\in C_{W}[{\cal A}_{i}]$ for one of the ${\cal A}_{i}$'s. We will therefore assume for the rest of this section that ${\cal A}$ is connected, with the understanding that when it is not, we are working with one of its connected components.\\

\noindent \textit{Lemma 2:} Let $p$ be a bulk point in the causal wedge of $\partial M$. Then $p\in C_{W}[{\cal A}]$ if and only if both of the following hold:
\begin{align}&C^{+}(p)\cap D_{\partial}[{\cal A}]\neq \varnothing \\
& C^{-}(p)\cap D_{\partial}[{\cal A}]\neq \varnothing.
\end{align}
\begin{proof}
\noindent (1) If $p\in C_{W}[{\cal A}]$, then $C^{\pm}(p)\cap D_{\partial}[{\cal A}]\neq \varnothing$.
By assumption, there are both past- and future-directed causal bulk curves from $p$ to $D_{\partial}[{\cal A}]$. By AdS hyperbolicity, future-directed bulk curves from $p$ cannot intersect or come arbitrarily close to intersecting past-directed bulk curves from $p$. So $X\equiv J^{\pm}(p)\cap D_{\partial}[{\cal A}]\neq \varnothing$ and $D_{\partial}[{\cal A}]-X$ is also not empty: i.e. there is a nonempty subset of $D_{\partial}[{\cal A}]$ which receives no causal curves from $p$. This immediately implies $\partial J^{\pm}(p)\cap D_{\partial}[{\cal A}]\neq \varnothing$, and therefore $C^{\pm}(p)\cap D_{\partial}[{\cal A}]\neq \varnothing$. \\
(2) If $C^{\pm}(p)\cap D_{\partial}[{\cal A}]\neq \varnothing$, then $p\in C_{W}[{\cal A}]$.
This is trivial: by assumption there exist both past- and future-directed causal curves from $D_{\partial}[{\cal A}]$, so $p\in C_{W}[{\cal A}]$. 

\end{proof}

The two lemmata above are sufficient for the construction of a proof relating the depth between two points and the causal wedges containing them. The theorems below are presented separately: the first applicable when one point is deeper than another globally (i.e. relative to the entire boundary), and the second applicable when one point is deeper than another relative to a particular boundary subregion. The second theorem may be thought of as a generalization of the first, but we present and prove them separately for pedagogical reasons.\\

\noindent \textit{Global Causal Wedge Inclusion:} Let $p$ and $q$ be two spacelike-separated bulk points. $p$ is deeper in the bulk if and only if any causal wedge containing $p$ also contains $q$.
\begin{proof} (1) If $p$ is deeper in the bulk than $q$, then any causal wedge containing $p$ also contains $q$. \\
Let $p\in C_{W}[{\cal A}]$, where ${\cal A}$ is a codimension 1 spacelike (acausal) surface, which we take to be connected by lemma 1. By the Gao-Wald theorem~\cite{GaoWal00}\footnote{Strictly speaking, the Gao-Wald theorem is a statement that bulk curves experience positive gravitational time delay relative to the boundary when the null generic condition is assumed. Away from such an assumption, a weaker version of the Gao-Wald theorem holds, which states that bulk curves experience a nonnegative gravitational time delay relative to boundary curves. As we do not assume the generic condition, we make use only of the weaker version of the theorem, which is sufficient for our proof.}, there exists a causal boundary curve from every point $x^{-}\in C^{-}(p)$ to every point $x^{+}\in C^{+}(p)$. Define $A^{\pm}(p)\equiv C^{\pm}(p)\cap D_{\partial}[{\cal A}]$, which is nonempty by lemma 2. This immediately implies that there exists a causal boundary curve from every point on $A^{+}(p)$ to every point on $A^{-}(p)$. Let $\gamma$ be one such  causal curve from $A^{+}(p)$ to $A^{-}(p)$; then $\gamma \in D_{\partial}[{\cal A}]$. Because $C^{\pm}(q)$ are sandwiched by $C^{\pm}(p)$, any causal curve from $C^{+}(p)$ to $C^{-}(p)$ must intersect both $C^{+}(q)$ and $C^{-}(q)$. So $C^{\pm}(q)\cap D_{\partial}[{\cal A}]\neq \varnothing$. By lemma 2, $q\in C_{W}[{\cal A}]$. This proves one direction. \\
(2) If any causal wedge containing $p$ also contains $q$, then $p$ is deeper in the bulk than $q$. \\
By contradiction. Suppose that for any ${\cal A}\in \partial M$, $p\in C_{W}[{\cal A}]$ implies $q\in C_{W}[{\partial A}]$, but there exists a boundary subregion on which $C^{\pm}(p)$ do not sandwich $C^{\pm}(q)$. Then there exists a boundary causal curve $\gamma$ from $x^{+}\in C^{+}(p)$ to $x^{-}\in C^{-}(p)$, where $\gamma$ does not intersect at least one of the cuts of $q$. Define a boundary domain of dependence using the $x^{\pm}$:
\begin{equation} D_{\partial}[{\cal A}]\equiv J^{+}(x^{-})\cap J^{-}(x^{+}).
\end{equation}
By construction, $p\in C_{W}[{\cal A}]$ ( $p$ actually lies on the causal surface of ${\cal A}$). So $q\in C_{W}[{\cal A}]$ as well. Again by construction, $\gamma$ does not intersect at least one of $C^{\pm}(q)$. Because the $C^{\pm}(q)$ are complete spatial slices (by assumption $q$ is not a boundary point), they can each at most intersect $D_{\partial}[{\cal A}]$ on a spatial slice (or, in the degenerate case, on $x^{\pm}$). However, since $\gamma$ does not intersect at least one of $C^{\pm}(q)$ anywhere, $C^{\pm}(q)\cap D_{\partial}[{\cal A}]=\varnothing$. By lemma 2, $q\notin C_{W}[{\cal A}]$, and we have arrived at a contradiction.
\end{proof}

%APPLICATIONS TO SURFACES

This theorem agrees with the intuitive idea that the existence of null geodesics that do not reach $\partial M$ may be necessary for a realization of the sandwich cut configuration; put differently, the existence of bulk regions which can only be causally accessed by nesting larger boundary regions is likely to require that some null geodesics never reach $\partial M$\footnote{I thank S. Fischetti for calling my attention to this point.}.

The relative version of the theorem may be proven with few modifications:\\

\noindent \textit{Relative Causal Wedge Inclusion:} Let $p$ and $q$ be spacelike separated bulk points in the causal wedge of $\partial M$. Let ${\cal A}\subset \partial M$ be an acausal, closed, $(d-1)$-dimensional subregion. $p$ is deeper than $q$ relative to ${\cal A}$ if and only if for any ${\cal A'}$ such that $D_{\partial}[{\cal A'}]\subset D_{\partial} [{\cal A}]$, $p\in C_{W}[{\cal A}'] \ \Rightarrow q\in C_{W}[{\cal A}']$.

\begin{proof}
(1) If $p$ is deeper than $q$ relative to ${\cal A}$, then for any ${\cal A'}$ as above,  $p\in C_{W}[{\cal A}'] \ \Rightarrow q\in C_{W}[{\cal A}']$. \\
The proof follows that of the global theorem \textit{mutatis mutandis}. Let $p\in C_{W}[{\cal A}']$, and define ${\cal A}^{' \pm}(p)$ as above. By definition there exists a causal curve $\gamma$ from $A^{'+}(p)$ to $A^{'-}(p)$ on $D_{\partial}[{\cal A}']$. Therefore, on $D_{\partial}[{\cal A}]$, the $C^{\pm}(q)$ are sandwiched by the $C^{\pm}(p)$. By the same logic as above, $\gamma$ intersects $C^{\pm}(q)$ on ${\cal A}$, but since $\gamma\subset D_{\partial}^{+}[{\cal A}']$, we find that $\gamma$ intersects $C^{\pm}(q)$ on $D_{\partial}^{+}[{\cal A}']$, which immediately implies that $q\in C_{W}[{\cal A}']$. 
(2) If for any ${\cal A'}$, $p\in C_{W}[{\cal A}'] \ \Rightarrow q\in C_{W}[{\cal A}']$, then $p$ is deeper than $q$ relative to ${\cal A}$.\\
By contradiction. By assumption there exists a causal curve $\gamma$ on $D_{\partial}[{\cal A}]$ from $C^{+}(p)$ to $C^{-}(p)$ such that $\gamma$ does not intersect $C^{\pm}(q)$. Consider now the domain of dependence $D_{\partial}[{\cal A}']$ constructed by bringing $\gamma$'s endpoints slightly into $\gamma$. Then $D_{\partial}[{\cal A}']\subset D_{\partial}[{\cal A}]$, but $C^{\pm}(q)\cap D_{\partial}[{\cal A}']=\varnothing$ by the reasoning in the proof of the global theorem. So $ q\notin C_{W}[{\cal A}']$: a contradiction. 
\end{proof}

\section{Discussion}\label{sec:dis}

We have given a covariant definition qualifying comparative depth of bulk points in terms of lightcone cuts, and more generally in terms of causal access to the boundary. This partial ordering of bulk points accomodates depth perception both relative to the entire boundary, and relative to a particularly boundary subregion. The definition in question also reduces to depth as measured by a usual notion of a standard holographic coordinate in pure AdS and static AdS black holes with isotropic spatial slices. Efficient curves, which locally minimize travel distance deeper in the bulk, can be used (when they exist) as a measure of comparative bulk depth. Finally, the intuition in which larger subregions on the boundary should correspond to deeper bulk regions is realized in our definition in the context of the causal wedge.

This definition of bulk depth relates deeper bulk points to (1) the singularity structure of $(d+3)$-point Lorentzian correlators at longer time-separation, and (2) the nesting structure of causal wedges at progressively larger distance scales. We therefore gain an understanding of how points deep in the bulk are sensitive to infrared phenomena in the dual field theory. Since flowing from the UV to the IR is considered a coarse-graining procedure, points deep in the bulk can be viewed as related to boundary points via a coarse-graining mechanism (e.g.~\cite{BalKra99}). This was realized precisely in entanglement renormalization tensor network schemes, starting with~\cite{Swi09}, on constant time slices of AdS$_{3}$. On a speculative level, it is interesting to ask if our qualification of one bulk point being deeper than another can be used in a similar way to construct more general bulk tensor networks, with bulk depth corresponding to coarse-graining. 

%In particular, if the lightcone cuts of two points are sandwiched, perhaps the points may be related via an isometric tensor in some continuous bulk tensor network, while points with crossing cuts are at the same entanglement scale or related by unitary tensors. 

On a less speculative level, it would be interesting to further develop the connection between nonlocal observables and our local definition of bulk depth. In particular, what is the relation between the entanglement wedges containing points at different depths? When do boundary-anchored efficient curves exist, and when are they geodesics?  Does the length of boundary-anchored efficient curves have information theoretic interpretation in the dual field theory?

Other interesting directions include an extension of our definition to the entire boundary domain of influence. While there is no procedure that yields these lightcone cuts from field theory data, we may ask how to appropriately define bulk depth under the assumption that the lightcone cuts have been recovered in some way. Finally, an event horizon is sufficient for the existence of sandwiched cuts, and therefore for a global definition of bulk depth to apply; it would be valuable to also understand the necessary conditions for the sandwich cut configuration to exist. 
 
We close with a comment on quantum corrections: since bulk-point singularities are robust against perturbative $1/N$ corrections, we expect that the definition of bulk depth given here is valid to Planck-sized neighborhoods in perturbatively quantum bulk spacetimes. As we have only assumed the Achronal Averaged Null Curvature Condition, we expect that our proofs on the causal wedge inclusion as well as the existence of the sandwich configuration will hold under inclusion of perturbative quantum effects as well.

\end{spacing}

\section*{Acknowlegements}
It is a pleasure to thank R. Bousso, J. Cano, B. Czech, D. Engelhardt, S. Fischetti, D. Harlow, G. Horowitz, H. Verlinde,  A. Wall, and C. Zukowski for helpful discussions. 
This work was supported in part by NSF grant PHY-1620059.

\bibliographystyle{JHEP}

\bibliography{all}

\end{document}